# Rare correlated sequences detection with DSSSD detector in a real-time mode at the Dubna gas-filled recoil separator

Yu.S.Tsyganov , A.N.Polyakov

tyra@jinr.ru

FLNR, JINR 141980 Dubna, Moscow region, Russian Federation

## Abstract

Method of active correlations is a key one to suppress radically beam associated backgrounds in heavy ion induced complete fusion nuclear reactions. This property is enhanced significantly when applying DSSSD detector instead of position sensitive resistive strip detector. Example of application in the $^{240}Pu+^{48}Ca \rightarrow 114+3n$ reaction is presented. An edge effect between neighbor strips is considered too. The value of the mentioned effect is measured as 15.9 % for ER-α correlation chains. **C++ Builder GFS-2016** program package for spectral analysis is presented. Calibration parameters (totally 352) were obtained from $^{nat}Yb+^{48}Ca$ nuclear reaction. New ultra intense FLNR (JINR) DC-280 cyclotron that is expected to be put into operation in the nearest future is also discussed from the viewpoint of the described method developments.

## 1. Introduction

About fifty years ago, it was theoretically predicted that unknown elements with numbers of neutrons (N) equal to184, and number of protons 114, 120, or 126 could form what is called an island of stability. Their half-lives to radioactive decay could be much, much longer than those of elements that have 2 to 10 protons less and especially those with 8 to 20 less neutrons. The half-lives of the nuclei with N=184 and Z=104-108 could be so long that they could even be comparable to the age of our Earth ($5 \cdot 10^9$ years) [1].

The scientists at the Dubna Gas Filled Recoil Separator DGFRS) of the Flerov Laboratory of Nuclear Reactions at the Joint Institute for Nuclear Research in Russia developed a new approach in which neutron-rich, long-lived radioactive targets from neptunium to californium were bombarded with very neutron-rich, double magic calcium-48 to make nuclei with neutron numbers 170-177 , much closer to 184. From 1999 to 2015 they reported the discoveries of new elements with Z=113,114,115,116,117,118 [2].

## 2. Detection module of the DGFRS: present status.

In long-term experiments aimed to the synthesis of Super Heavy Elements (SHE) one should take into account that yield of the products under investigation is small enough, usually – one per days – one per month, thus the role of the detection system and focal plane detector is quite significant as well as beam intensity requirements. Since 2015, to increase the position granularity of the detectors, which reduces the probability of observing sequences of random events that could imitate decay chains of synthesized nuclei, the new focal-plane detector has been used. It consists of 120x60 cm$^2$ 48x128 strips Micron Semiconductor **D**ouble **S**ide **S**ilicon **S**trip **D**etector (**DSSSD**, type **A39-45**). Design of this detector is shown in the Fig.1. The detection system of the DGFRS was calibrated by registering the recoil nuclei and decays (α, SF) of known isotopes of No and Th and their descendants produced in the reactions $^{206}Pb(^{48}Ca,2n)$ and $^{nat}Yb(^{48}Ca, 3-5n)$, respectively. Before implantation into the focal plane detector, the separated ERs passed through a time-of-flight (TOF) measuring system that consists of two (start and stop) multiwire proportional chambers filled with pentane at ≈1.6 Torr [3,4]. The TOF system allows to distinguishing recoils coming from the separator and passing through the TOF system from signals, arising from α decay or SF of the implanted nuclei (without TOF of ΔE$_1$ or ΔE$_2$ signals). In order to eliminate the background from the fast light charged particles (protons, α's, etc produced from direct reactions of projectiles with the DGFRS media) with signal amplitudes lower than the registration threshold of the TOF detector, a "VETO" silicon detector is placed behind the front detector.

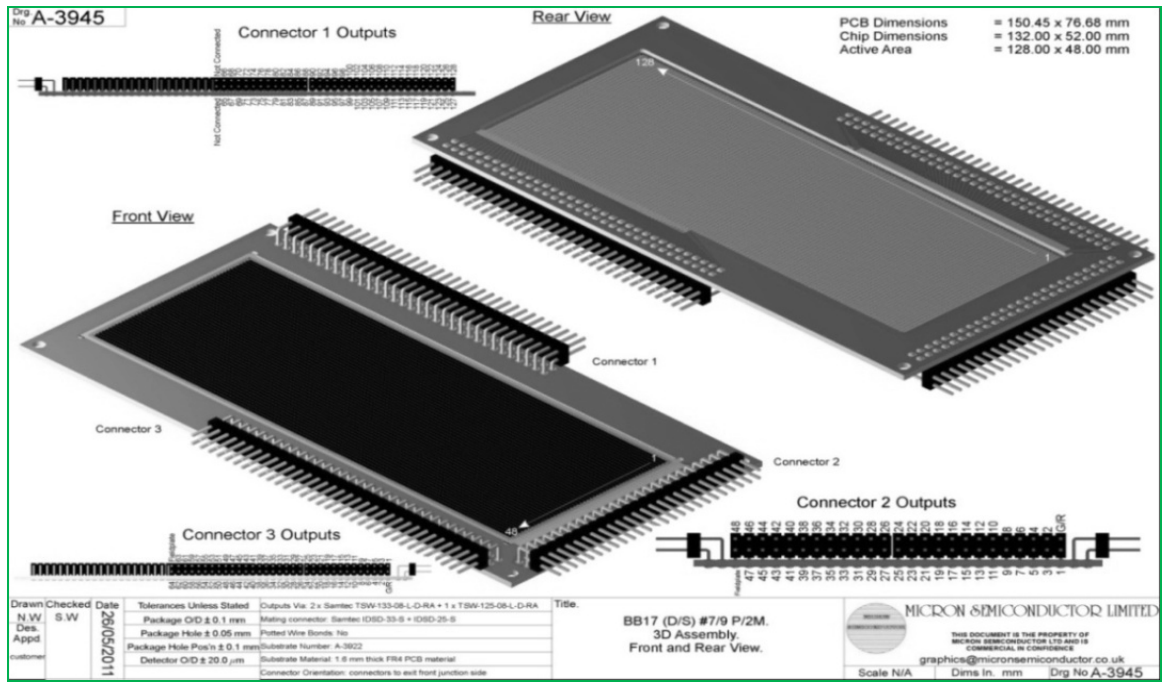

**Fig.1** Design of **A-395** Micron Semiconductor focal plane detector of the DGFRS. Active area =128.0 x 48.0 mm, PCB dimensions = 150.45 x 76.55 mm, Chip Dimension=132.0 x 52.0 mm.

From theoretical calculations and the available experimental data, one can estimate the expected α-particle energies of the ERs and their descendant nuclei that could be produced in a specific heavy-ion induced reaction of synthesis. For α particles emitted by the parent or daughter nuclei, it is possible to chose wide enough energy and time gates $\Delta E_{\alpha 1}$, $\Delta t_{\alpha 1}$, $\Delta E_{\alpha 2}$, $\Delta t_{\alpha 2}$, etc. and to employ a special low-background detection scheme –method " *active correlations*"

**3. Method of "active correlations"**

The simple, but very effective idea of the mentioned method is as following. PC-based Builder C++ program is aimed at searching in real-time mode of time-energy-position recoil-alpha links, using the two matrix representation of the DSSSD detector separately for ER matrix and α-matrix. In each case of "alpha particle" signal detection, a comparison with "recoil"-matrix is made. If the elapsed time difference between "recoil" and "alpha particle" within preset time value, the system turns on the cyclotron beam chopper which deflects the heavy ion beam in the injection line of the U-400 FLNR cyclotron for a definite time interval (usually 0.5-2 min).

The next step of the computer code ignores horizontal position (128 strips from p-n junction side) of the forthcoming alpha-particle signal during the beam-off interval. If such decay takes place in the same vertical strip (48 strips) that generated the pause, the duration of the beam-off interval is prolonged by a factor 5-10. The dead time of the system, associated with interrupting the beam is about 110 μs, including linear growth chopper operation delay(~10 μs) and estimated heavy ion orbit life-time(~60 μS). In contrast to former resistive layer PIPS detector applications [5-11], using of DSSSD detector one has three main specific features:

1. ER matrix (48x128 elements) de-facto already exists due to discrete composition of the DSSSD detector;

2. Om the other hand, edge effects between the neighbor p-n junction side strips should be taken into account (128 back strips in our case);

3. From the viewpoint of radiation durability of the DSSSD it should be mentioned that detector is operated strongly in total depletion mode.

New version of software, reported below, takes into account points1, 2. As to point 3, there is seemingly no radical method to eliminate the problem, except for conservative one. That is, for example, to change the electromagnetic separator design, to use additional degraders before focal plane detector, etc.

**3. GNS-2016 Builder C++ program package**

GNS-2016 Builder C++ program package has been designed to work together with new DSSSD based detection module of DGFRS and appropriate electronics. It consists of two main parts:

-ERAS-2016.exe –data taker and file writer also used to generate beam stop signal;

-MONITOR_2016.exe – a visualization unit also used for exact tuning of TOF-ΔE low pressure, pentane filled module;

- Some programs used for testing electronics modules are also within this package.

## 3.1 ERAS-2016 C++ data taker code

**ERAS-2016** C++ program (***ER** –**A**lpha **S**equence*) is designed to provide data taking, file writing and to search for ***ER-α*** correlated sequences in a real-time mode. It provides event-by-event data taking from the **DGFRS** detection system. The block-diagram and of this process and the flow chart of the code are shown in the Fig.2a,b, respectively. Note, that beam is chopped in the cyclotron injection line, when the value of $^{48}$Ca projectile energy is small enough ($^{48}$Ca beam energy is ~18 kV at the position of the beam chopper). The code branch, that is responsible for real-time search for ER-α sequences is shown in gray in Fig.2b.

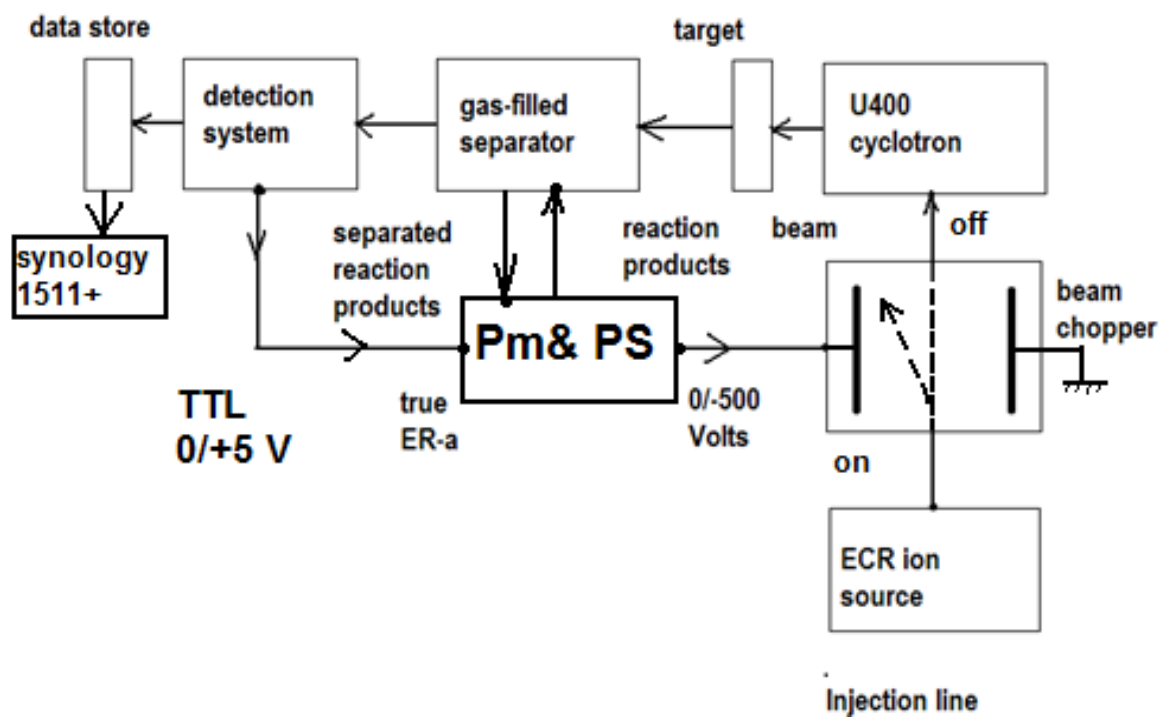

**Fig.2a** Block-diagram of providing beam stop by using ERACS-2016 C++code. ***Pm & PS** – **P**arameter **m**onitoring and **P**rotection System of the DGFRS (see ref. below)

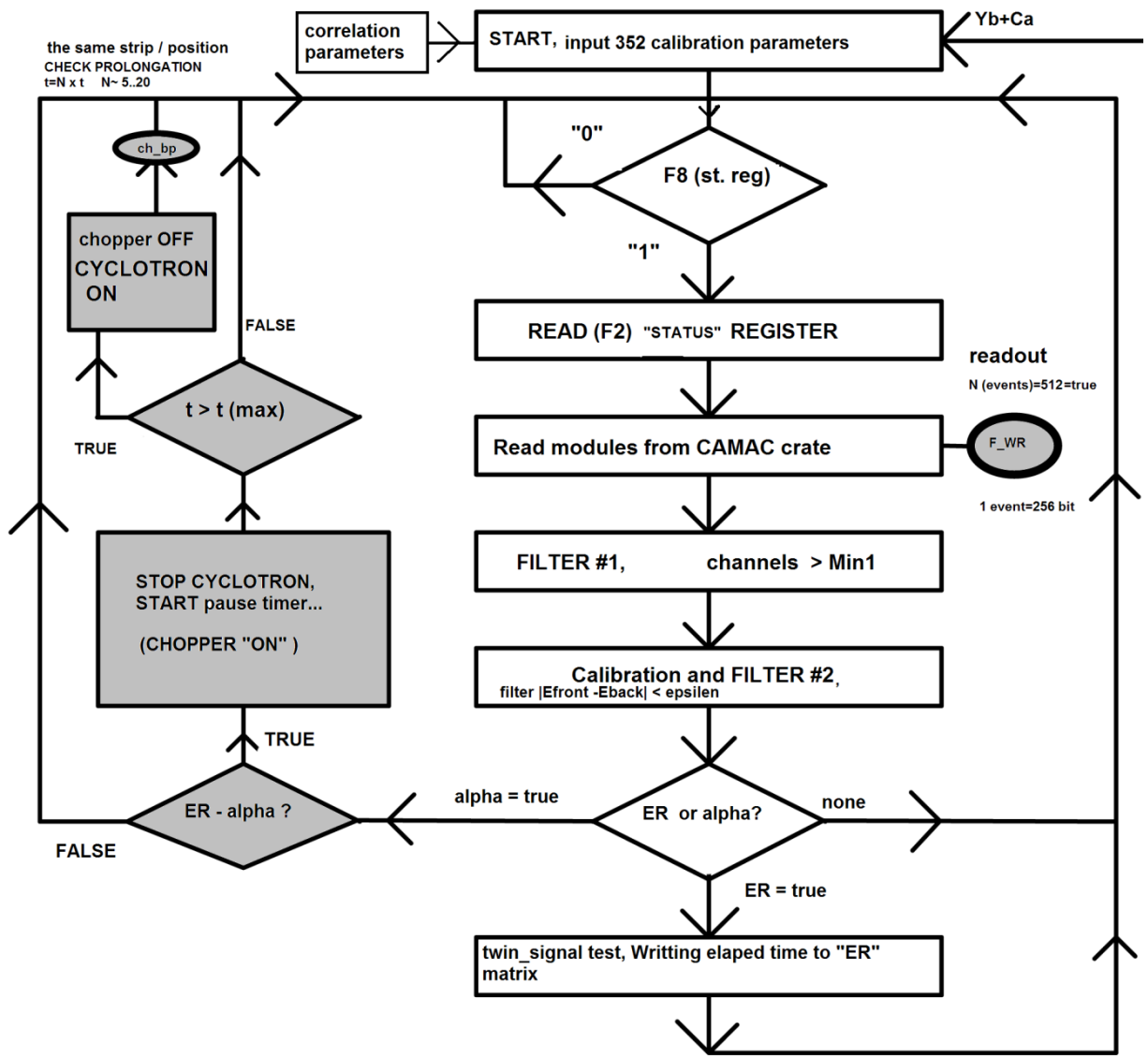

**Fig.2b** The flowchart of the process of search for ER-alpha chains. $E_f$-value of front strip energy signal, $E_b$-value of back strip energy signal. (k=5-20, usually). CH_BP – **ch**eck **b**eam **p**rolongation routine, F_WR – **f**ile **wr**iting.

ERAS correlation parameter list is represented below:

- ER-α correlation time to provide a beam stop (this can be a function of incoming α-particle energy);
- Integer value (5-20) which denotes that in the case of prolongation a beam off interval, the pause will be a factor 5 to 20 longer;

- Minimum and maximum values of ER and α-particle signals set to stop the beam and min and max value of alpha particle signal set for prolongation of beam-off interval;
- Minimum and maximum values of TOF and $\Delta E_{1,2}$ signals measured with low pressure gaseous TOF module.

Routines "Filter#1" and "Filter#2" shown in Fig.2b provide filtering of incoming signals according to channel number and energy, respectively. The routine "check prolongation" is active only when the bean chopper is in "switch on" state, otherwise it provides no extra operation. Distinguishing between true/false of a Boolean variable "*twin signal*" is performed by reading of the appropriate eight bits of "status" register. Each bit in "1" state corresponds to operation of one of 16-input analog-to-digital converters. Additionally, ERAS program generates text file with parameters of every beam stop. It includes energy signals of recoil and alpha particle from both front and back strip, elapsed time of the ER signal and time difference between alpha particle signal and ER (recoil) signal, numbers of the strips and one bit marker (0/1) indicating simultaneous operation of two neighboring strips on *p-n* junction side. In $^{251,249}$Cf+$^{48}$Ca reaction at beam intensity ~0.7 pμA, such "double" events from the DSSSD back side strips amounted to 15.9% of total number (618 events of 3867). The last effect is demonstrated in Fig.3 that shows test of status register with an external α-particle source.

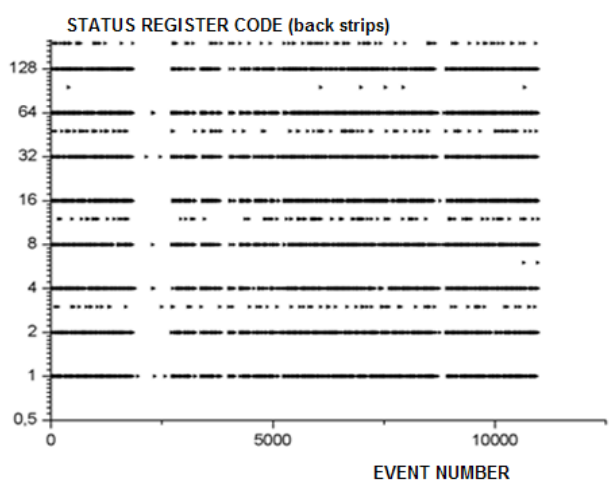

**Fig.3** Codes of the "status" register (see **Fig.2b**) for eight 16-input ADC's to measure back strip signals

Note, that reading a number that is a clear power of "2" means that only one back side strip signal is measured, whereas a mixed value (e.g. $2^0 + 2^1$ ...etc) denotes that signals from two neighbor strips are registered.

In the latter case, the program calculates actual back strip energy in the form:

$E_{back} = a_i \cdot N_i + b_i + a_{i+/-1} \cdot N_{i+/-1} + b_{i+/-1}$. Here, $(a_i, b_i)$ – calibration constants, i=1..128. An additional filtering condition should be fulfilled, that is: $|E^j_{front} - E^i_{back}| \leq \varepsilon$. Here $E_{front}$ is the energy signal value measured with front side strip and $\varepsilon$ is a small value preset in the program (~100 KeV) and (j,i) – front/back pixel indexes (48x128). Respectively, if ER signal is measured, two ER matrix elements are filled, one with i- number, another - with i+1 or i-1. Six points in Fig.3 located between ordinate tics "64" and "128" correspond to simultaneous operation of three (or more) back strips. Their fraction in the whole data array (~ 6/11000 = 0.05%) is, without doubt, negligible. In the DGFRS experiments such events are excluded from search for ER-α-α...SF sequences. Below, a fragment of C++ code of filling ER matrix elements is presented.

```
if (twin_signal==true )
    {
      if (strip_1 >=0 && strip_1 < 128)
    {
      MATRIX[strip][strip_1]=t_elapsed;   RECOIL[strip][strip_1]=energy;}
      if(strip_2 >=0 && strip_2 < 128)
      {MATRIX[strip][strip_2]=t_elapsed;   RECOIL[strip][strip_2]=energy;}
    }
```

Here, MATRIX[i][j] – ER matrix and its matrix element is *t_elapsed* – the elapsed time value of the event obtained from a standard CAMAC electronics, *twin_signal* – is a boolean variable (case of "true" corresponds two strips operation), *strip_1,2* – numbers of actual back side (p-n junction side) strips. RECOIL[i][j] – a supplementary matrix for recoil energy values.

If α-like signal is detected and *twin_signal* value is equal to *true,* the actual back strip number index is defined as (C++ coding):

*strip_act = (CHANNEL1 > CHANNEL2) ? strip1 : strip2.*

Here, *CHANNEL1* and *CHANNEL2* correspond to the channels read from the first and second back side strips, respectively. Therefore, when the code makes the choice whether to make a breakpoint or not in case of *twin_signal* is *true*, the schematics shown in Fig.4a is realized. In this example, the measured channel number in back strip j is greater than one for j+1. In the opposite case, when *N (i,j) < N(i,j+1),* beam pause will be provided through comparison of elapsed time of α-like signal with ER matrix element (i, j+1).

To support this approach in case of *twin_signal = true,* Fig.4b shows two-dimensional histogram for two neighbor back strips energies under condition $|E^j_{front}-E^i_{back}| \leq \varepsilon$ and for the energy value of *9261 KeV (within +/- 100 KeV* interval) measured by front detector. The calibration line is from α-decay of $^{217}$Th produced in the reaction $^{nat}Yb+^{48}Ca \rightarrow {}^{217}Th+3n.$

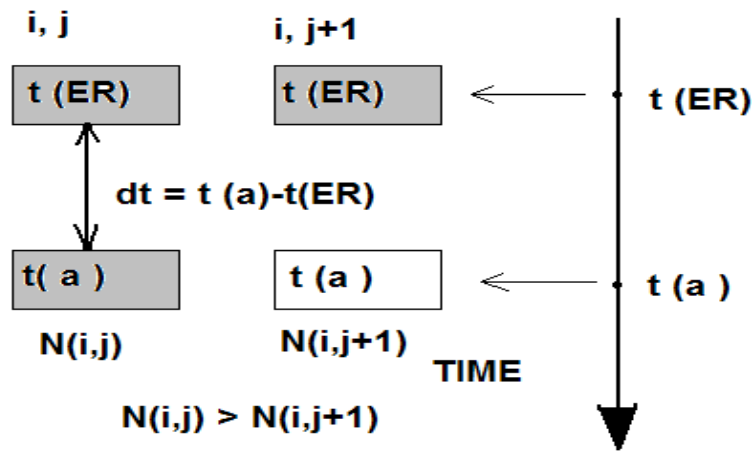

**Fig.4a** Schematics for beam stop process if *twin_signal = true. N(i,j)* – channel for back strip number j and *N(i,j+1)* – the same for j+1. I is a front strip number, *t (ER)* is an element of ER matrix for appropriate indexes, *t(a)* – elapsed time value of alpha-like signal.

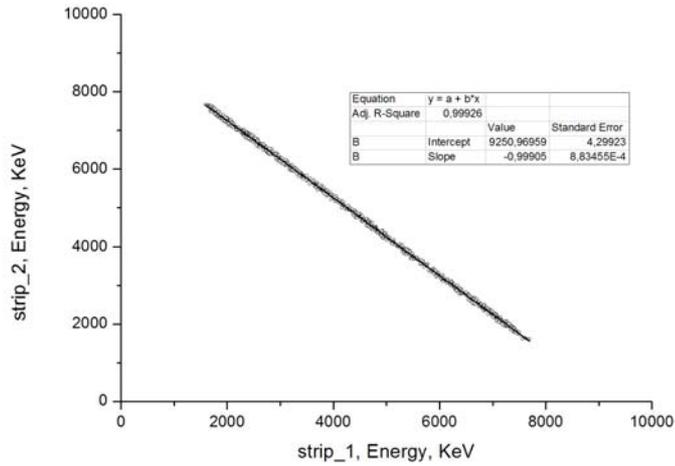

Fig.4b Two dimensional histogram for neighbor strips energy values for $^{217}$Th isotope alpha decay, *if twin_signal=true*

## 3.2 Monitor-2016 C++ code for file processing

C++ Builder Monitor-2016 program is designed for processing of files generated by ERAS. In Fig.5a,b the main window of this application is shown. The program constructs spectra for each front and back strip and for ΔE and TOF signals (totally, 250 histograms).

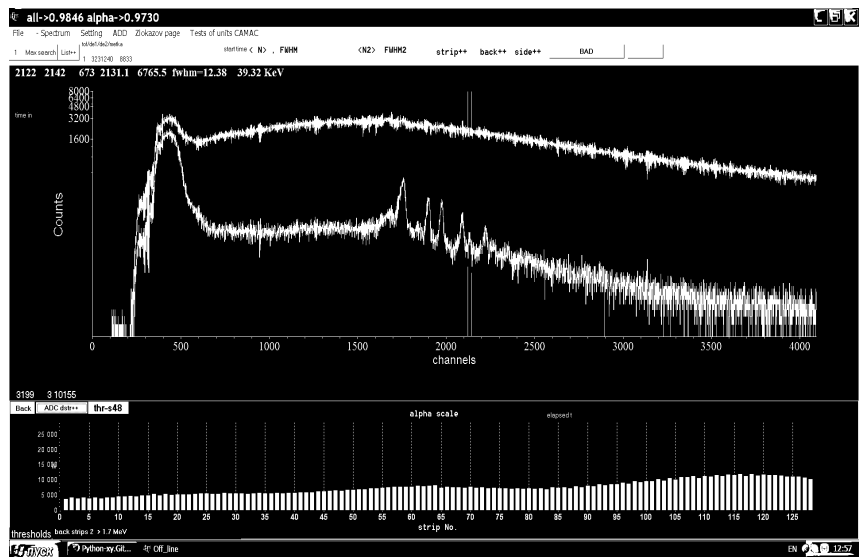

**Fig.5a** The main form of Monitor-2016 application (PtrScr). Bottom histogram – distribution over to 128 back side strips. Two upper histograms represent recoil and alpha-particle signals. ($^{251,249}$Cf+$^{48}$Ca reaction. //Fragment).

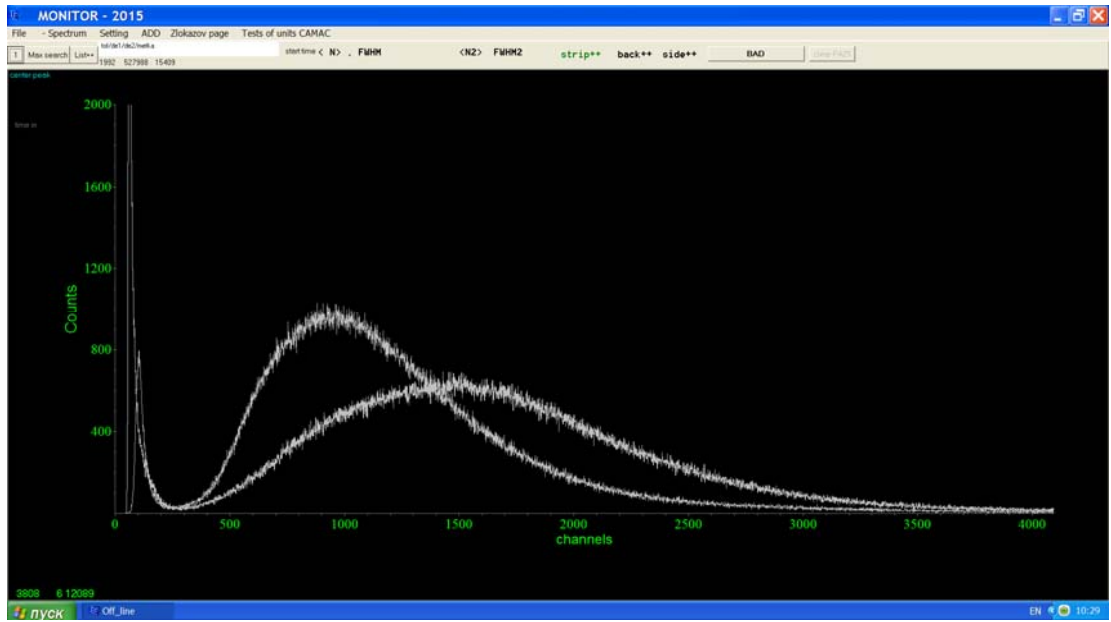

**Fig.5b** Typical ΔE spectra from START/STOP gaseous detectors. Left hands part (~ from 50 to 200 channels) is an alpha particle like spectra.

Except for building histograms, some specific spectra are built by the program. For example, it provides output files constructed as sum alpha spectra meeting a condition:

*a*) All signals TOF=0, $\Delta E_{1,2}=0$;

b) The same, but additionally, single-bit flight marker is equal to zero.

This flight marker is generated if at least one signal from start or stop gaseous counter exceeds a 40-mV threshold of a one-shot unit; in this case, the latter generates 0/+5 Volt output TTL signal with duration about 20 μs (preamplifier response to typical ER signal is ~0.5-1 Volt and about 50 mV for 5.5 MeV α-particles). Of course, with low-threshold one-shot unit, certain precautions must be made to avoid extra suppression of true α-particle signals. A special interactive VMRIA program package [12] was used to check this. In Fig.6 two α-like spectra are shown. To start processing those peaks VMRIA program should be placed to the same directory as MONITOR-2016 one. Then, loading of VMRIA is performed by pressing on of the MONITOR-2016 menu items just after the creation of an appropriate spectrum file with scaling of 20 KeV/channel from the MONITOR-2016. In this procedure, the first peak position approximations are marked manually in an interactive mode by experimentalists. The result of processing is shown in Table.1. Evidently, using of flight marker provides additional possibility to extract α-particle spectra. With applying this technique, a 99.6-99.7% suppression factor was achieved for an energy interval of 9–11.5 MeV in $^{251,249}$Cf+$^{48}$Ca reaction. It means, here only 0.3-0.4% of events can be considered alpha-like imitator signals. In the right-upper part of Fig.6, four peaks processed by VMRIA code are shown. Dots denote measured data and line corresponds to the results obtained from VMRIA processing. Both spectra correspond each other perfectly.

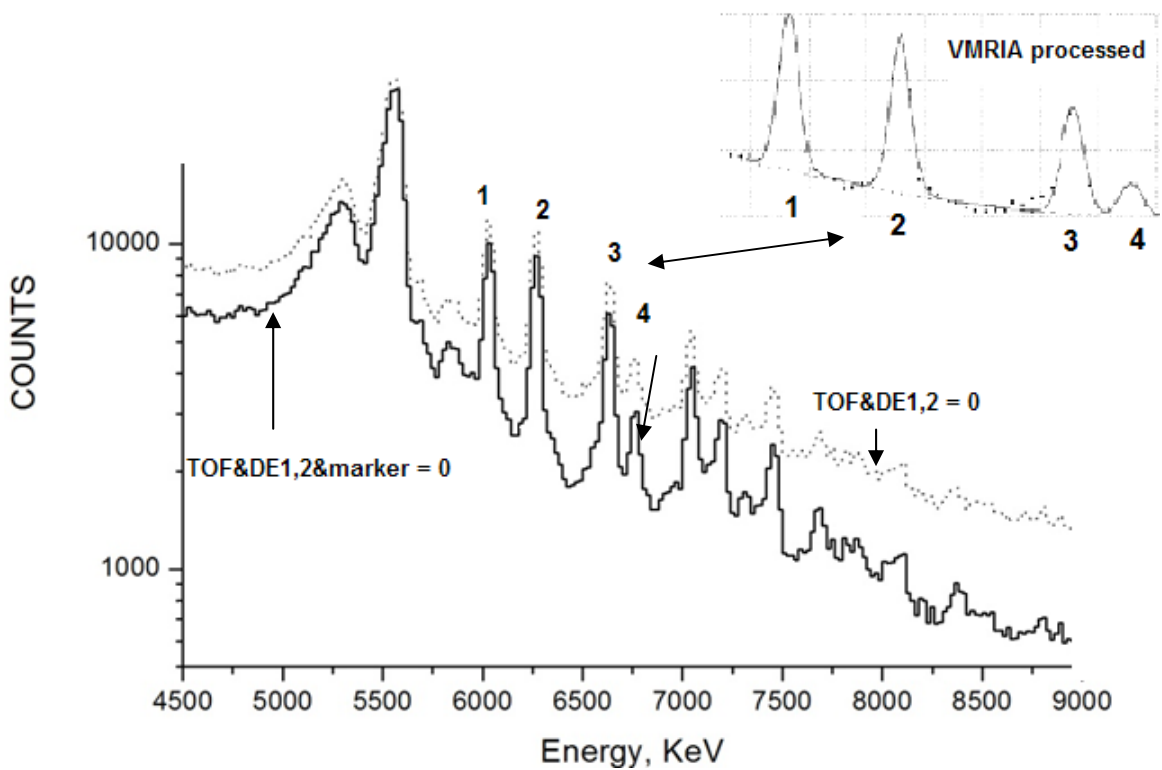

**Fig.6** Two spectra of α-like signals. Dotted line – selection under condition that every of three signals (TOF, ΔE1 and ΔE$_2$) is zero; solid line - the same, with single-bit flight marker required to be zero. Nuclear reaction is $^{251}$Cf+$^{48}$Ca. Four processed peaks (solid line from original spectrum) are marked from 1 to 4 in both histograms.

| Peak position (dotted, Fig.5, channels) | Peak position(solid) | Sum of peak (dotted) | Sum of peak(solid) |
|---|---|---|---|
| 301.99 | 302.00 | 19929 +/- 274 | 19600 +/- 303 |
| 313.89 | 313.89 | 19248 +/- 244 | 19279 +/- 280 |
| 332.16 | 332.19 | 13128 +/- 194 | 13051 +/- 225 |
| 338.49 | 338.42 | 4128 +/- 125 | 3929 +/- 155 |

**Table 1.** Comparison of the sum of peak and it's position for two spectra from Fig.5 after VMRIA code processing. Scaling factor is equal to 20 KeV/ch. (*Four processed peaks are indicated by numbers 1 to 4 in Fig.6.*)

## 4. Example of application of Builder C++ ERAS code in the $^{240}$Pu+$^{48}$Ca →Fl* reaction

In the long-term $^{240}$Pu+$^{48}$Ca→Fl* experiment, the beam was interrupted after the detection of a recoil signal with the expected implantation energy for Z=114 evaporation residues followed by an α-like signal in the front detector with the energy 9.8-11.5 MeV, in the same (or neighbor) DSSSD pixel. The ER energy interval was chosen to be 6 – 16 MeV. The triggering ER-α time interval was set to 1 s. The beam-off interval was set to 1 min. In this time interval, if an α-particle with $E_α$ =8.5 to 11.5 MeV was registered in the same front strip as the ER signal, the beam-off interval was automatically extended to 5 min. During the experiment, two chains were detected that were attributed to Z=114 nuclei [13]. These are presented in the Fig.7. Signals detected during the beam of interval are marked by shadow. Both chains of Z=114 are in fact pixel-to-pixel events, that is, all decays occur only in the one front and one back strips. With the measured value of edge effect of 15.9% the expected probability to register at least one event with *twin_signal=true* is about 0.32. In the nearest future we shall continue to study this effect more thoroughly. As to both ER signals 12.557 Mev and 10.195 MeV, the mean value of 11.376 MeV corresponds well to the systematic of Ref. [14,15] where this was formulated for the DGFRS detection module as:

*<E>≈25.9 – 0.125·Z =11.65 [MeV].*

These registered ER's amplitudes are also in a good agreement with Monte Carlo simulated spectrum for Z=114 from ref. [6] shown in Fig.8.

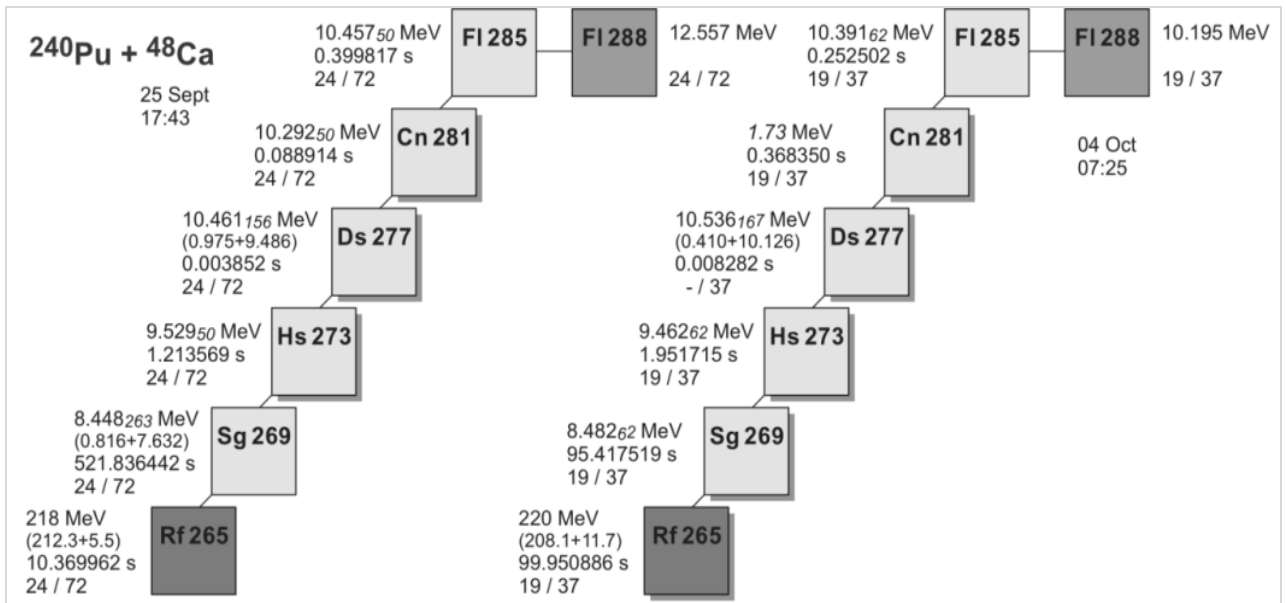

**Fig.7** Two decays chains of Z=114 nuclei detected in a real-time mode with DSSSD detector using ERAS code. (Shadows mark beam off interval)

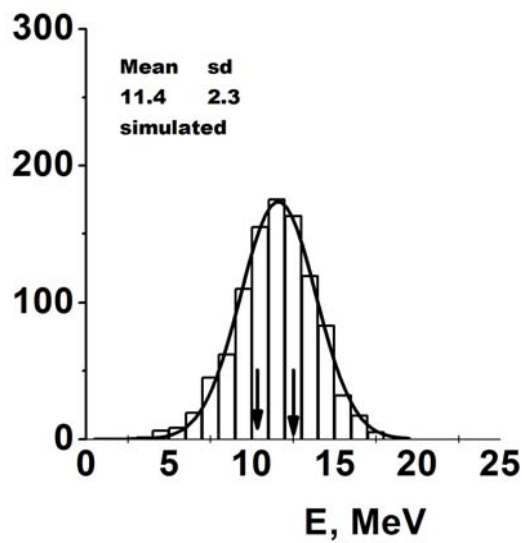

**Fig.8** Simulated Z=114 ER's spectrum. Two registered ER signals in a real-time mode are shown by arrows. Solid line is a Gaussian fit of simulated data.

## 5. Prospects: DC-280 high-intensity FLNR cyclotron project

During last years method of "active correlations" was extensively and successfully used in the long-term experiments aimed at the synthesis of new superheavy nuclei. It was used both with resistive layer PIPS detector and, since 2013, with large area Micron Semiconductor made DSSSD detector. This method was applied by extraction of ER-α correlated sequences in a real-time mode to switch off the intense $^{48}$Ca projectile beam. Of course, beam stops definitely mean losses of beam time. With a heavy ion beam intensity of ~1pμA, typical losses are about 0.5-10% that is quite acceptable for experimentalists. On the other hand, when FLNR DC-280 cyclotron (super-heavy element factory) will put into operation [16,17] and intensities of 5 to 10 pμA will be available, the requirements for real-time algorithms will be higher. Of course, one way to solve this problem is in the new design of the gas-filled separator, e.g. the greater bending angle etc. As to the algorithm itself, authors assume it reasonable to introduce more chains in consideration. One could use second (third…) alpha-signal sequence, use first ER signal as a triggering signal for shorter beam off interval in parallel to a standard ER-α, use more sophisticated TOF system and so on . Together with a future application of DSSSD detector with higher granularity, one should take into account the increase of edge effect between neighbor p-n junction side strips when developing of the data taking computer code.

As a final author's remark: because the present detection system of the DGFRS is in fact a part of a more general integrated system that includes the system of monitoring parameters and protection of the DGFRS [18], the latter is assumed to be upgraded in parallel in the nearest future, as well. The main trend is unification of the whole system.

## 6. Summary

Together with the higher granularity advantage, using of DSSSD detector arises some local problems, the edge effect between neighbor strips on p-n junction being one of them. With Borland's Builder C++ GFS-2016 program package this problem was solved. For ER-alpha sequences a value of 15.9% was measured as the quantity estimate of this effect at the DGFRS. The "active correlations" method was successfully applied in the $^{240}$Pu+$^{48}$Ca →Fl* complete fusion nuclear reaction. Measured by the DGFRS DSSSD detector, average ER's energy is in good agreement with the value calculated from the systematic. For our future projects, associated with putting into operation in 2017 of a new FLNR high-intensity DC-280 cyclotron, we plan to develop more sophisticated algorithms for searching only ER-α sequences in a real-time mode. Although a DSSSD radiation stability problem is outside the scope the present paper, authors note that it is actual. Indeed, in contrast to resistive layer PIPS detector, DSSSD is operated only in full depletion mode and in the future project one should pay more attention to this problem. An upgrade of the DGFRS parameter monitoring and protection system will go in parallel.

This paper is supported in part by the RFBR Grant № 16-52-55002/16. The authors are indebted to Dr.'s A.Voinov and M.Shumeiko for their help and fruitful discussions.